\documentclass[prb,twocolumn,showpacs,superscriptaddress]{revtex4}

\usepackage{graphicx}
\usepackage[centertags]{amsmath}

\newcommand{\cm}{\ensuremath{\,\mbox{cm}^{-1}}}
\newcommand{\K}{\ensuremath{\,\mbox{K}}}
\newcommand{\celsius}{\ensuremath{\,{}^\circ}\!C}

\newcommand{\Tc}{T$_{c}$}
\newcommand{\Tn}{T$_{N}$}

\begin{document}

\title{Infrared and magnetic characterization of the multiferroic Bi$_{2}$FeCrO$_{6}$ thin
films in a broad temperature range}

\author{ S.~Kamba}\email{kamba@fzu.cz} \affiliation{Institute of Physics ASCR,
v.v.i. Na Slovance~2, 182 21 Prague~8, Czech Republic}
\author{D.~Nuzhnyy}
\affiliation{Institute of Physics ASCR, v.v.i. Na Slovance~2, 182 21 Prague~8, Czech
Republic}
\author{R. Nechache}
\affiliation{INRS - \'{E}nergie, Materiaux et T\'{e}l\'{e}communications, 1650, boulevard
Lionel-Boulet, Varennes (Quebec), J3X 1S2, Canada}
\author{K. Z\'{a}v\v{e}ta}
\affiliation{Faculty of Mathematics and Physics, Charles University, V
Hole\v{s}ovi\v{c}k\'{a}ch 2, 18000 Prague 8, Czech Republic }
\author{D. Ni\v{z}\v{n}ansk\'{y}} \affiliation{Faculty of Natural Sciences,
Charles University, Albertov 6, 128 43 Prague 2, and Institute of Inorganic Chemistry,
v.v.i., 250 68 Rez near Prague, Czech Republic}
\author{E. \v{S}antav\'{a}}
\affiliation{Institute of Physics ASCR, v.v.i. Na Slovance~2, 182 21 Prague~8, Czech
Republic}
\author{C. Harnagea}
\affiliation{INRS - \'{E}nergie, Materiaux et T\'{e}l\'{e}communications, 1650, boulevard
Lionel-Boulet, Varennes (Quebec), J3X 1S2, Canada}
\author{A. Pignolet}
\affiliation{INRS - \'{E}nergie, Materiaux et T\'{e}l\'{e}communications, 1650, boulevard
Lionel-Boulet, Varennes (Quebec), J3X 1S2, Canada}

\date{\today}


\begin{abstract}

Infrared reflectance spectra of an epitaxial Bi$_{2}$FeCrO$_{6}$ thin film prepared by
pulsed laser deposition on LaAlO$_{3}$ substrate were recorded between 10 and 900\K. No
evidence for a phase transition to the paraelectric phase was observed, but some phonon
anomalies were revealed near 600\K. Most of the polar modes exhibit only a gradual
softening, which results in a continuous increase of the static permittivity on heating.
It indicates that the ferroelectric phase transition should occur somewhere above 900\K.
Magnetic measurements performed up to 1000\K\, revealed a possible magnetic phase
transition between 600 and 800\K, but the exact critical temperature cannot be determined
due to a strong diamagnetic signal from the substrate. Nevertheless, our experimental
data show that the B-site ordered Bi$_{2}$FeCrO$_{6}$ is one of the rare high-temperature
multiferroics.

\end{abstract}

\maketitle

\section{Introduction}

Magnetoelectric multiferroic materials which exhibit simultaneously ferroelectric and
magnetic order are promising for new generation of random access memories (RAM), where
the information can be written by electric field and read non-destructively by magnetic
sensing. Such memories avoid the weak points of the ferroelectric RAMs (destructive
reading causes fatigue) as well as of magnetic RAMs (high electric current is needed for
overwriting, which rules out high integration of magnetic RAMs). Unfortunately, there are
not many magnetoelectric multiferroics known up to now and only a few of them have both
magnetic and ferroelectric critical temperatures above room temperature. Therefore there
is nowadays an intensive search for magnetoelectric multiferroic materials with high
magnetization and spontaneous polarization above room
temperature.\cite{fiebig05,cheong07}

Baettig and Spaldin predicted from \textit{ab initio} calculations that the chemically
ordered double perovskite Bi$_{2}$FeCrO$_{6}$ (BFCO) will have - at zero temperature - a
polarization of $\sim$80\,$\mu$C/cm$^{2}$ and a magnetization of $\sim$160 emu/cm$^{3}$
(2 $\mu_{B}$ per formula unit).\cite{baettig05a} Such properties far exceed the
properties of any known multiferroic. Nechache \textit{et al.} for the first time
experimentally prepared epitaxial thin film of BFCO which exhibited at room temperature
(RT) a polarization of 2.8\,$\mu$C/cm$^{2}$ and a saturated magnetization of
0.26$\mu_{B}$ per unit cell.\cite{nechache06,nechache07} Recently Kim et al.\cite{kim07}
reported a remanent polarization of 60\,$\mu$C/cm$^{2}$ at 77\K, for their
BiFe$_{0.5}$Cr$_{0.5}$O$_{3}$ solid solution epitaxial filmsa nd Alexe even measured
70-80\,$\mu$C/cm$^{2}$ at RT on the films grown by Nechache et al.\cite{alexe07} The
magnetic \Tn\, and ferroelectric \Tc\, phase transition temperatures in ordered BFCO are
not known up to now. Beattig \textit{et al.}\cite{baettig05b} predicted from first
principles a N\'{e}el temperature \Tn\, near 100\K, which was not confirmed in the
experiments performed by Nechache \textit{et al.}, who observed magnetic order at RT.
Very recently Suchomel \textit{et al.}\cite{suchomel07} prepared BFCO ceramics and
observed a magnetic phase transition below 130\K, but it is worth noting that their
ceramics exhibited chemical disorder of the Fe$^{3+}$ and Cr$^{3+}$ cations on the
perovskite B site, which reduces \Tn. Thus it is not excluded that \Tn\, can be higher in
ordered samples as in the case for the ordered BFCO films reported by Nechache at
al.\cite{nechache07b}

Determination of the ferroelectric phase transition temperature is not possible from the
low-frequency dielectric measurements due to the too high DC conductivity of the BFCO
film. The extrinsic leakage conductivity does not play an appreciable role in the THz
dielectric response of the film, therefore high-frequency dielectric studies are
advantageous. For this purpose we performed infrared (IR) measurements including
investigation in hardly achievable far infrared (FIR) range below 200\cm, which can give
information about the complete phonon contributions to the static permittivity (note that
in the case of displacive ferroelectrics only polar phonons are responsible for the
dielectric anomaly near T$_{c}$). Moreover, IR spectra usually change at the
ferroelectric (structural) phase transition temperature due to the change of selection
rules for IR active polar phonons. Therefore the IR spectra (including FIR) of the BFCO
film can help to estimate its \Tc\, as well as the symmetry of the high-temperature
phase.

IR studies of ferroelectric thin films are rather rare in the literature and up to now
almost only FIR transmission spectra of the films deposited on FIR-transparent substrates
like Si, sapphire or MgO were investigated. FIR transmission can give results only in a
limited frequency range determined by the transparency window of the substrates, which is
mostly very narrow, particularly at high temperatures (e.g. sapphire is partially
transparent only below 150\cm\, at 900\K).\cite{kamba05} IR reflectance can yield results
in a much broader spectral range, but its sensitivity is limited a) by the thickness of
the film, b) by the strengths of polar phonons in the IR spectra and c) by the IR
properties of the substrate. Our experience shows that the substrates with buffer
electrodes are not suitable due to the negative permittivity of the buffer layers, which
reduces the sensitivity of the method. Therefore, dielectric substrates, which do not
show any strongly temperature-dependent IR reflectivity spectra, are the most suitable
for reflectance studies of thin films. Nevertheless, IR reflectance spectra of the thin
films deposited on the substrate are strongly influenced by the substrate, since the thin
films are partially transparent for the IR wavelength. Therefore both IR spectra of the
bare substrate and of the film on the substrate should be measured at the same
temperatures and the film properties are evaluated from the spectra fits to such a
multilayer system. This method was used only twice in the literature for room or
low-temperature IR studies of SrTiO$_{3}$ films.\cite{almeida06,yamada06} Here, we will
use this method for the first time above room temperature and up to 900\K.

In this paper we shall show that the static permittivity of the BFCO thin film,
determined from the polar phonon contributions, increases monotonically on heating to
900\K, due to the phonon softening. Some phonon anomalies, probably connected with a
magnetic phase transition, were observed near 600\K, but no dramatic changes, such as
those usually related to a ferroelectric phase transition, were observed. Therefore, it
seems that the phase transition to the paraelectric phase in BFCO thin film occurs above
the highest investigated temperature of 900\K. We shall also report on our study of the
magnetic properties of the BFCO film up to 1000\K\, and we shall show that the magnetic
phase transition occurs between 600 and 800\K.

\section{Experimental}
BFCO films were grown directly on (100)-oriented SrTiO$_{3}$ substrates doped with 0.5
wt\% of Nb (abbreviated STO:Nb) as well as on a (100)-oriented LaAlO$_{3}$ substrate,
better suited for IR measurements. An epitaxial 210 nm thick film deposited on the
conducting STO:Nb substrate was used for M\"{o}ssbauer spectroscopy and vibrating sample
magnetometry studies. XRD and TEM data have shown that the thin film is epitaxial. Weak
superlattice spots showed evidence of partial chemical order of the Fe and Cr
cations.\cite{nechache07b}

The SrTiO$_{3}$ substrate exhibits a strongly temperature-dependent FIR spectra due to
the presence of an optic soft mode, which gives large inaccuracies in evaluation of the
FIR properties of the thin film. We therefore studied FIR spectra of the BFCO film
deposited on non-conducting LaAlO$_{3}$ substrates with the size 5x10x0.5\,mm. Since
polar phonons are rather weak in BFCO, we investigated a 600\,nm thick film. The relaxed
film on LaAlO$_{3}$ substrate (orientation (001)) was epitaxial with orientation (100),
but it revealed only very weak superlattice spots, so its chemical order in the B site
was only partial. XRD analysis of the film on LaAlO$_{3}$ revealed a slight Bi-deficiency
in the BFCO phase as well as about 5\% of Cr-doped Bi$_{2}$O$_{3}$ secondary phase, which
were not observed for the films on SrTiO$_{3}$.

The unpolarized FIR and IR reflectance spectra were taken using a Bruker IFS 113v FTIR
spectrometer at temperatures between 10 and 900\K\, with the resolution of 2\cm. An
optistat CF cryostat from Oxford Instruments equipped with polyethylene windows was used
for cooling the sample down to 10\K, while a commercial high temperature cell SPECAC P/N
5850 was used for heating it up to 900\K. A helium-cooled Si bolometer operating at
1.6\K\, was used as a detector at low temperature measurements, while pyroelectric DTGS
detectors were used for the IR measurements above RT.

Magnetic properties of the BFCO films on substrates 3x3 mm in size were investigated
using a PPMS 14 vibrating-sample magnetometer (Quantum design) between 3 and 1000\K.

The M\"{o}ssbauer spectrum measurement was carried out in the Conversion electron
M\"{o}ssbauer spectroscopy (CEMS) mode with $^{57}$Co diffused into an Rh matrix as a
source moving with a constant acceleration. The spectrum was accumulated for 7 days.
Classical M\"{o}ssbauer spectroscopy in transmission mode could not be used due to the
small volume of the investigated thin films. The Wissel spectrometer was calibrated by
means of a standard $\alpha$-Fe foil, and the isomer shift was expressed with respect to
this standard at 293\K. The fitting of the spectra was performed using the NORMOS
program. The CEMS method requires a conducting sample, therefore we investigated the thin
film deposited on a conducting STO:Nb substrate, while the films deposited on
non-conducting LaAlO$_{3}$ were more suitable for the IR studies.

\section{Results and discussion}

\begin{figure}
  \begin{center}
    \includegraphics[width=80mm]{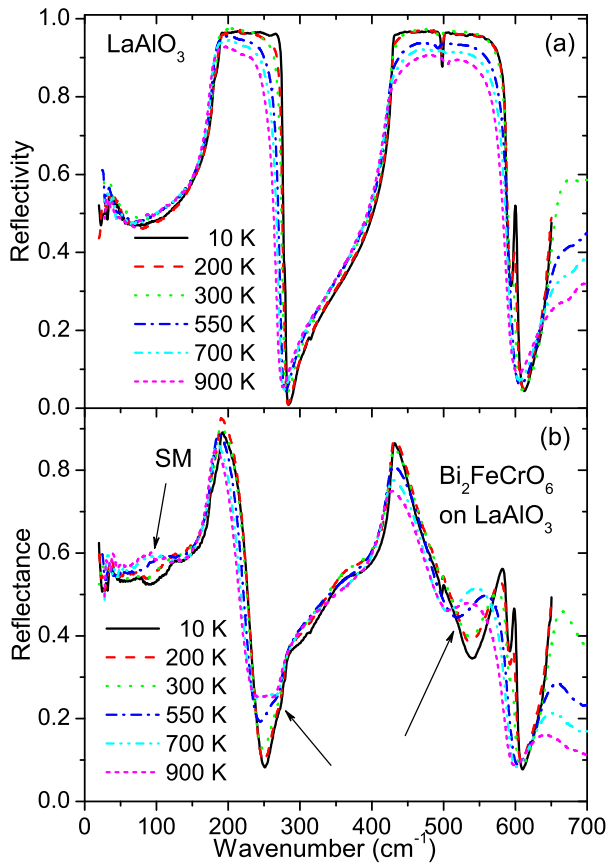}
  \end{center}
    \caption{(color online) (a) Infrared reflectivity spectra of  LaAlO$_{3}$ substrate and (b) reflectance spectra
    of Bi$_{2}$FeCrO$_{6}$
    film (600 nm thick) deposited on LaAlO$_{3}$. The arrows show anomalous phonons. Note that
    1\cm\, corresponds to 30\,GHz.}
    \label{Fig1}
\end{figure}

Fig.~\ref{Fig1} shows IR reflectance spectra of both a pure LaAlO$_{3}$ substrate and a
BFCO thin film (deposited on LaAlO$_{3}$) at selected temperatures between 10 and 900\K.
Only small temperature dependence of the reflectivity spectra of the LaAlO$_{3}$
substrate can be seen, mostly due to an increase in phonon damping with temperature. Also
the sharp peaks near 500 and 600\cm\, gradually disappear due to a second order
structural phase transition in LaAlO$_{3}$ from trigonal to cubic phase at
800\K.\cite{mueller68}

\begin{figure}
  \begin{center}
    \includegraphics[width=80mm]{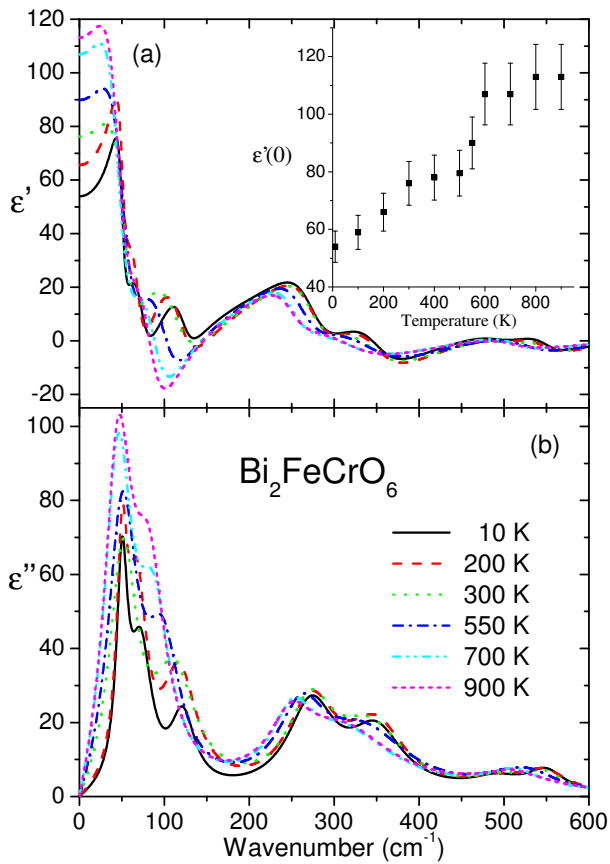}
  \end{center}
    \caption{(color online) Complex dielectric spectra of the BFCO film obtained at selected temperatures
    from the fit of the reflectance spectra in Fig.~\ref{Fig1}. Frequencies of the peaks
    in $\varepsilon''$($\omega$) spectra roughly correspond
to the TO phonon frequencies.
    Note the continuous increase of the static permittivity $\varepsilon'$(0) with rising temperature (see inset).}
    \label{Fig2}
\end{figure}

IR reflectance spectra of the semitransparent BFCO film deposited on the opaque
LaAlO$_{3}$ substrate exhibits more pronounced changes with temperature mainly near 100,
250 and 550\cm\, (marked by arrows in Fig.~\ref{Fig1}). For the detailed analysis we
evaluated the complex permittivity
$\varepsilon^{*}(\omega)=\varepsilon'(\omega)-i\varepsilon''(\omega)$ spectra of the film
(see Fig.\ref{Fig2}) using the following procedure: The reflectivity R($\omega$) of the
bare substrate at each temperature was first fitted using the formula
\begin{equation}\label{refl}
R(\omega)=\left|\frac{\sqrt{\varepsilon^{*}(\omega)}-1}{\sqrt{\varepsilon^{*}(\omega)}+1}\right|^2
,\end{equation}

where for the $\varepsilon^{*}$($\omega$) the factorized form of the complex permittivity
\cite{gervais83} was used
\begin{equation}\label{eps4p}
\varepsilon^{*}(\omega)=\varepsilon_{\infty}\prod_{j}\frac{\omega^{2}_{LOj}-\omega^{2}+i\omega\gamma_{LOj}}{\omega^{2}_{TOj}-\omega^{2}+i\omega\gamma_{TOj}}\,.
\end{equation}
$\omega_{TOj}$ and $\omega_{LOj}$ denote the frequencies of the j-th transverse and
longitudinal polar phonon, respectively, and $\gamma$$_{TOj}$ and $\gamma$$_{LOj}$ denote
their corresponding damping constants. The high-frequency permittivity
$\varepsilon_{\infty}$ results from the electron absorption processes and from the phonon
contribution above 600\cm. Then the spectrum of the two-slab system (film + substrate)
was fitted using the full formula for the coherent reflectance of a two-layer
system\cite{Born,Heavens} where the oscillator parameters of the substrate were fixed, in
order to determine the oscillator parameters of the polar phonons in the film. We note
that for the film IR spectra fits we used the classical Lorentz model of the damped
harmonic oscillators instead of Eq.~(\ref{eps4p})

\begin{equation}
\label{eps3p}
 \varepsilon^*(\omega)
 = \varepsilon_{\infty} + \sum_{j=1}^{n}
\frac{\Delta\varepsilon_{j}\omega_{TOj}^{2}} {\omega_{TOj}^{2} -
\omega^2+\textrm{i}\omega\gamma_{TOj}} \, ,
\end{equation}
where $\Delta\varepsilon_{j}$ means the contribution of the j-th mode to the static
permittivity. The rest of the parameters in Eq.~(\ref{eps3p}) have the same meaning as in
Eq.~(\ref{eps4p}). Eq.~(\ref{eps4p}) is more suitable for the reflectivity fits of phonon
spectra with a large TO-LO splitting, when both kinds of phonon modes have different
damping. Such a model was necessary to use for a good fit of the LaAlO$_{3}$ substrate.
Eq.~(\ref{eps3p}) is more appropriate for fitting of the reflectivity spectra of phonons
with a small TO-LO splitting and/or transmission spectra which do not show up anomalies
at LO frequencies. It has fewer parameters and gives acceptable physical results, while
the former model can sometimes yield un-physical negative dielectric losses, when the
parameters are not properly chosen.

\begin{figure}
  \begin{center}
    \includegraphics[width=80mm]{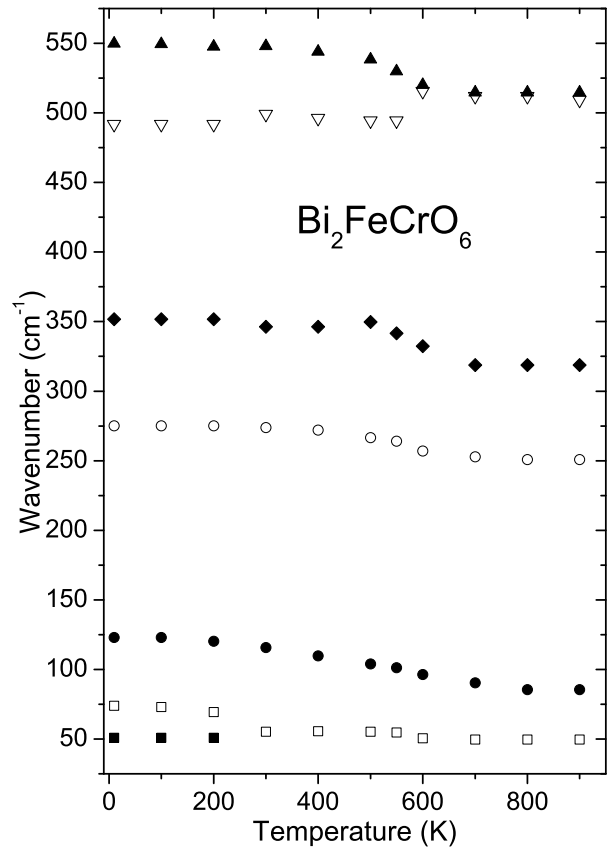}
  \end{center}
    \caption{Temperature dependence of the transverse polar phonon eigen-frequencies in the
    BFCO film deposited on LaAlO$_{3}$.}
    \label{Fig3}
\end{figure}

Complex dielectric spectra of BFCO film obtained from the above described fit of the IR
reflectance spectra displayed on Fig.~\ref{Fig1} are plotted in Fig.~\ref{Fig2}. The
temperature dependence of TO phonon frequencies is plotted in Fig.~\ref{Fig3}. One can
clearly see the shift of most of the phonon modes to lower frequencies on heating (phonon
softening). It causes the gradual increase of the static permittivity with rising
temperature (see inset in Fig.~\ref{Fig2}).

Six modes (respectively seven below 200\K) were necessary for the fits. Let us compare
the number of observed polar modes with the prediction of factor group analysis. BFCO
crystallizes in the rhombohedral space group $R3-C^{4}_{3}$. Bi, Fe and Cr cations have
the site symmetry $C_{3}$(1), while oxygen ions have the $C_{1}$ site
symmetry.\cite{baettig05a,baettig05b} Factor group analysis of the lattice vibrations
based on tables published by Rousseau et al. \cite{rousseau81} yields the following optic
phonons
\begin{eqnarray}
\Gamma_{R3} = 9A(z,x^{2}+y^{2},z^{2}) + 9 E(x,y,x^{2}-y^{2},xy,xz,yz).
 \label{eq:rhombo}
\end{eqnarray}
It means that the 9$A$ and 9$E$ modes are both Raman and IR active. The analysis gives
also additional 1$A$+1$E$ acoustic modes. The modes with the $A$ symmetry are active in
spectra with the electric vector \textbf{E} of the IR wave parallel to the $z$ axis,
while the $E$ modes are active in \textbf{E}$\parallel$ $x,y$ spectra. The rest of
symbols ($z^{2}$, $xy$ etc.) in brackets of the group analysis in Eq.~\ref{eq:rhombo}
shows components of Raman tensors, in which the phonons are Raman active. Our epitaxial
BFCO film is (001) oriented and since we measure the in-plane response, we see mostly the
$E$ symmetry modes in our FIR spectra. We resolved 7 modes in the low-temperature spectra
although 9 $E$ modes are allowed. This is quite reasonable, if we take into account that
some of the modes have small intensity or they may overlap with other modes. It is worth
to note that the TO phonon frequencies in BFCO correspond very well to the $E$ symmetry
TO phonon frequencies in chemically and structurally related
BiFeO$_{3}$.\cite{kamba07,hermet07,lobo07}

One mode near 50\cm\, disappears from the FIR spectra above 200\K\, (see black solid
squares in Fig.~\ref{Fig3}). Such change could be a hint of some structural phase
transition, but the FIR spectra near 50\cm\, are rather noisy at high temperatures, so we
cannot exclude that the mode is present in the spectra also at higher temperatures, but
we do not resolve it due to lower sensitivity of the high-temperature FIR experiment. The
absence of any other phonon anomalies at higher frequencies also does not give some phase
transition near 200\K.

Interesting phonon anomalies are seen near 600\K\, (see Fig.~\ref{Fig3}). Some mode
frequencies show relatively large temperature changes and splitting of the modes near 490
and 550\cm\, that almost disappears above 600\K. However, it is important to stress that
no mode from the doublet disappears above 600\K. Both modes remain in the spectra with
similar frequencies near 520\cm\, up to 900\K. Suchomel \textit{at al.},\cite{suchomel07}
observed decomposition of BFCO ceramics on heating above 400\celsius. This effect could
be also responsible for the phonon anomalies seen near 600\K\, in our film, but we have
to emphasize that we did not observe any decomposition in our sample (placed in a vacuum
chamber of the spectrometer), because the IR spectra and magnetic properties (see below)
were reproducible before and after the thermal cycling.

Phonon anomalies near \Tn=640\,K, similar to ours in Fig.~\ref{Fig3}, were observed in
Raman spectra of BiFeO$_{3}$ \cite{haumont06} and they were explained by spin-phonon
coupling. We will discuss this possibility below together with the magnetic data.

The phonon frequency changes seen near 600\K\, can be a consequence of some phase
transition, but probably not a ferroelectric one because we see a gradual increase of the
static permittivity $\varepsilon'$(0) (from phonon contributions) on heating (see inset
in Fig.~\ref{Fig2}), while for a ferroelectric transition a maximum in
$\varepsilon$'($T$) should be seen near \Tc. It seems that the ferroelectric phase
transition in BFCO lies above 900\K, like for BiFeO$_{3}$. Note that BiFeO$_{3}$ has a
rhombohedral $R3c$ structure with a structural phase transition (according to earlier
literature) to cubic paraelectric $Pm\bar{3}m$ phase near 1120\K. Very recent structural
studies\cite{scott07} revealed an intermediate orthorhombic phase with the space group
$C_{2v}^{1}-P2mm$ or $C_{2v}^{11}-C2mm$ at temperatures between $\sim$1100 and
$\sim$1200\K\, and probably only above $\sim$1200\K\, BiFeO$_{3}$ transforms into the
cubic and simultaneously conducting phase.\cite{scott07}

If we assume that the structural phase sequence in BFCO is the
same as in BiFeO$_{3}$, than the following factor group analysis
of the optic phonons applies in the orthorhombic phase
\begin{equation}
 \Gamma_{P2mm} = 10A_{1}(z,x^{2})+4A_{2}(xy)+7B_{1}(x,xz)+6B_{2}(y,yz) .
 \label{eq:ortho}
\end{equation}

It means that instead of the 18 modes in the rhombohedral
structure, 23 IR active modes (13 in \textbf{E}$\parallel$ $x,y$)
should be seen in the IR spectra of the orthorhombic phase.
Unfortunately, no new mode appears in our FIR spectra, so we do
not see any evidence for a phase transition into the orthorhombic
phase at temperatures below 900\K.

On the other hand, in cubic paraelectric phase the following optic
modes are expected:

\begin{multline}
  \label{eq:cubic}
  \mathrm{\Gamma_{Pm\bar{3}m} = 4 F_{1u}(x)  + 2 F_{2u}(-) + 1 A_{1g}(x^{2}+y^{2}+z^{2})} \\
  \mathrm{ + 1 E_{g}(x^{2}+y^{2}-2z^{2}, \sqrt{3}x^{2}-\sqrt{3}y^{2})} \\
  \mathrm{ + 2 F_{2g}(xy,yz,xz).}
\end{multline}

It seems that only 4 phonons of $F_{1u}$ symmetry should be seen in the FIR spectra and 4
modes ($A_{1g}$, $E_{g}$ and $F_{2g}$ symmetries) in Raman spectra. We see 6 modes in
Fig.~\ref{Fig3}. It means that BFCO probably remains in the rhombohedral phase in the
whole investigated temperature range up to 900\K\, and the structural phase transition
only occurs, similarly as in BiFeO$_{3}$, at higher temperatures. The absence of a phase
transition from the ferroelectric to paraelectric phase below 900\K\, is also supported
by observed gradual increase of the static permittivity on rising temperature - see inset
in Fig.~\ref{Fig2}. Further investigations, like high-temperature structural or second
harmonic generation, are needed for revealing the T$_{c}$ and symmetry of the
high-temperature phase(s).

\begin{figure}
  \begin{center}
    \includegraphics[width=85mm]{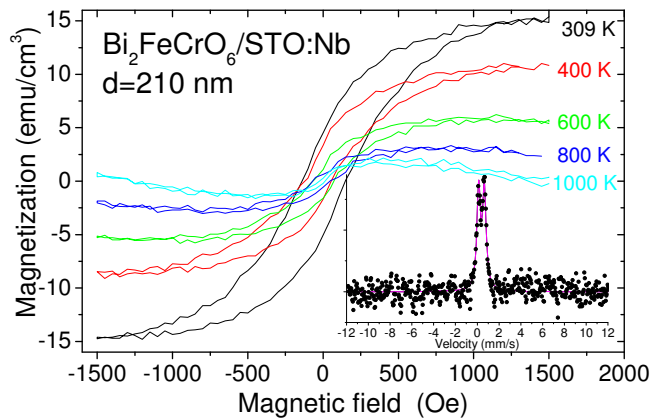}
  \end{center}
    \caption{(color online) Magnetic hysteresis loops of BFCO thin B-site ordered film deposited
    on STO:Nb substrate measured at selected temperatures up to 1000\K.
    Inset shows the room-temperature CEMS M\"{o}ssbauer spectrum of the same film together with its fit.}
    \label{Fig4}
\end{figure}

Let us discuss the magnetic properties of BFCO thin film, which we investigated by means
of vibration magnetometry and CEMS M\"{o}ssbauer spectroscopy. The BFCO film deposited on
the LaAlO$_{3}$ substrate, originally investigated by IR spectroscopy, exhibits a strong
diamagnetic signal from the substrate and the magnetic hysteresis loops were only
revealed at low temperatures below 20\K\, (not shown here). It can be explained by a weak
B-site order which suppresses the magnetic phase transition temperature. On the other
hand, the well-ordered BFCO thin film on the STO:Nb substrate exhibits nice magnetic
hysteresis loops not only at RT but also at higher temperatures (see Fig.~\ref{Fig4}).
Negative slope of magnetization at higher magnetic fields seen above 800\K\, can be
explained by a diamagnetic contribution of the STO:Nb substrate, but below 600\K\, the
open hysteresis loop is clearly seen. The value of saturated magnetization is typical for
antiferromagnets with weak ferromagnetism induced by a canted spin structure and the
value of RT spontaneous magnetization corresponds well to the previously published
results.\cite{nechache06,nechache07,nechache07b} The low value ($\sim$ 0.3
$\mu_{B}$/f.u.) of magnetization at saturation of the film regarding the expected
theoretical value of 2 $\mu_{B}$/f.u. \cite{baettig05a} could be explained by (i) the
Fe-Cr ordering which may be only partial, (ii) the partial chemical disorder that
generates an antiferromagnetic antisites contribution (Fe-Fe, Cr-Cr), or/and (iii) the
partial relaxation of the strain in the film leading to a more distorted structure.

\begin{figure}
  \begin{center}
    \includegraphics[width=80mm]{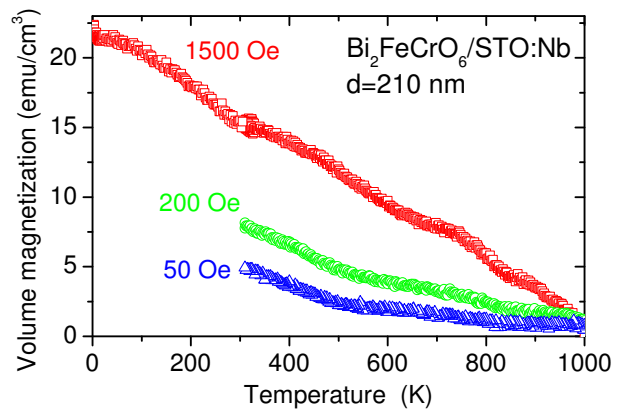}
  \end{center}
    \caption{(color online) Temperature dependence of the magnetization in BFCO/STO:Nb film at various magnetic fields.}
    \label{Fig5}
\end{figure}

Fig.~\ref{Fig5} shows the temperature dependence of the magnetization at various magnetic
fields. The magnetic experiments above and below 300\K\, were performed separately, which
is probably the reason of the change of slope seen at 300\K\, in temperature dependence
of magnetization measured at 1500\,Oe. The magnetization remains nonzero up to 1000\K\,
in the field of 1500\,Oe, but the hysteresis loop is very slim at temperatures above
800\K. It is difficult to determine exactly the magnetic phase transition from
Fig.~\ref{Fig4} and \ref{Fig5}, but it seems that it could lie somewhere between 600 and
800\K. It is worth noting that Beattig \textit{et al.}\cite{baettig05b} predicted $T_{N}$
in BFCO only near 100\K, which can correspond to recent result of Suchomel et al.
\cite{suchomel07}, who claim, based on transmission M\"{o}ssbauer spectrum of BFCO
ceramics, that $T_{N}$ lies below 130\K. Suchomel's low $T_{N}$ can be explained by
chemical disorder in the Fe and Cr cations in the [111] direction,\cite{suchomel07} while
our thin film is at least partially chemically ordered (we observed satellites peaks both
in the XRD and in the selected area electron diffraction patterns taken by TEM).

The inset of Fig.~\ref{Fig4} shows the room-temperature CEMS M\"{o}ssbauer spectrum of
the BFCO film. Surprisingly, only a doublet is seen, which is typical for the
\textit{paramagnetic} state of Fe$^{3+}$ ions in the sample, while a sextet is expected
in a magnetically ordered state. This is rather puzzling because clear magnetic
hysteresis loops are seen in the same sample by vibration magnetometry (Fig.~\ref{Fig4}).
Conversion Electron M\"{o}ssbauer spectroscopy of $^{57}$Fe is based on the detection of
electrons with the energy of 7.3 keV which were knocked out of the K shell of the
$^{57}$Fe atom after re-emission of the gamma quantum originally resonantly absorbed by
the $^{57}$Fe nucleus. The release is almost instantaneous (within 10$^{-7}$ s) and has a
rather high probability. Most of these electrons are again absorbed in the material, but
some of them, depending on the depth, where the electron emission occurs and on the
electron work function, reach the surface of the sample and are finally detected. The
depth from which the information is collected by the CEMS method is usually $\sim$ 200 nm
depending on the absorption properties of the material, the electrons emitted from deeper
regions of the sample do not reach the surface. Our thin film is only 210 nm thick, which
means that we should see the CEMS signal from the whole volume of the film. However, the
film is not only magnetic but also ferroelectric, and electric field in the ferroelectric
domains should substantially influence the work function of the electrons. The
ferroelectric domain structure of BFCO is complex and assuming it is similar to that
reported for BiFeO$_{3}$,\cite{chu07} the polarization (and related internal electric
field) is oriented 41.8$^{\circ}$ or even 131.8$^{\circ}$ to the normal surface of (001)
oriented thin film. Therefore, the emitted electrons are returned back to the film and
most of them lose their energy, are absorbed and do not leave the film. Only electrons
emitted from a very thin surface layer ($\sim$ 10 nm) may reach the surface and are
detected in the CEMS experiment. The thin film surface layer is most likely non-magnetic
(probably due to chemical disorder of Fe and Cr cations at the surface), therefore only a
doublet is observed in our CEMS M\"{o}ssbauer spectra shown in the inset of
Fig.~\ref{Fig4}, although the volume of the film is magnetically ordered, as seen from
the magnetic hysteresis loops measured by vibrating-sample magnetometry.

Nevetheless, we have to stress that we repeated the CEMS M\"{o}ssbauer experiment also
with another BFCO film (thickness 86 nm, STO:Nb substrate), which exhibited strong
satellites in the XRD (i.e. a higher Fe and Cr chemical order than in the previous
sample), as well as broad magnetic hysteresis loop and still we found only doublet in
CEMS spectra typical for the paramagnetic order. Finally, we note that the doublet cannot
originate from the substrate, because the M\"{o}ssbauer spectrum is sensitive only to the
Fe cations not present in the STO:Nb substrate.

\begin{table}
\caption{Comparison of the fit parameters of M\"{o}ssbauer spectra in Fig.~\ref{Fig1} and
in Ref.~\cite{suchomel07}.}
\begin{tabular}{|l l l l |}\hline
 & Our data& & Suchomel et al.\cite{suchomel07}\\
 &(CEMS) & & (Trans. mode) \\\hline
Isomer shift $\delta$ &0.39 mm/s& & 0.39 mm/s \\
Quadrupole splitting $\Delta E_{q}$ &0.52 mm/s & & 0.48 mm/s \\
"Peak width"-FWHM ($\Gamma$) &0.39 mm/s & & N/A\\
  \hline
\end{tabular}
\label{moesbauer}
\end{table}

Parameters of the M\"{o}ssbauer spectra fit are summarized in Table~\ref{moesbauer}. From
the Fe isomer shift $\delta$ the valency of the iron can be clearly estimated. The
Fe$^{3+}$ cations in the oxidic compound have its isomer shift in the range of 0.1-0.5
mm/s while the Fe$^{2+}$ cations show $\delta$ in the range of 0.8-1.5
mm/s.\cite{menil85} Our obtained value $\delta$=0.39\,mm/s confirms the absence of
Fe$^{2+}$ states and the presence of only Fe$^{3+}$ states in the investigated film.
M\"{o}ssbauer spectra also allow determining the site symmetry for Fe$^{3+}$ cations in
the structure. According to Refs.\cite{menil85,parmentier99}, the usual isomer shift
values for Fe$^{3+}$ in the case of the spectra measured at RT are as follows: 0.10-0.30
mm/s for Fe$^{3+}$ in a tetrahedral site, 0.28-0.50 mm/s for Fe$^{3+}$ in an octahedral
site. When we compare the above-mentioned ranges with our $\delta$=0.39 mm/s, we can
confirm that Fe$^{3+}$ in BFCO is in the octahedral position.

When we compare our fitting parameters in Table~\ref{moesbauer} with the parameters
obtained from the M\"{o}ssbauer spectra (measured in the transmission mode) of disordered
ceramics published by Suchomel \textit{at al.},\cite{suchomel07}, we can state that both
ceramics and surface layer of our thin film have the same or similar non-magnetic
structure at RT, although the magnetic measurements by vibrating-sample magnetometry give
evidence for a magnetic order in the thin film far above RT.

In the light of our high-temperature magnetic data we can suggest that the phonon
anomalies seen near 600\K\, are due to a magnetic phase transition. Near this temperature
a sudden drop of the permittivity is seen, which is typical for spin-phonon
coupling.\cite{smolenskii82} Nevertheless, further magnetic, structural and dielectrics
studies are necessary for the confirmation of this suggestion.

\section{conclusion}

The complex dielectric response of a BFCO film was determined by a novel method - IR
reflectance of a BFCO thin film deposited on a LaAlO$_{3}$ substrate. Most of the polar
phonons seen in the IR spectra reveal gradual softening on heating from 20 to 900\K,
which causes a progressive increase of the static permittivity with increasing
temperature. Therefore we speculate that the ferroelectric phase transition lies
(similarly to the related BiFeO$_{3}$) above 900\K, although some phonon anomalies,
probably connected with the magnetic phase transition, were observed near 600\K. Magnetic
properties of the BFCO thin film were investigated between 6 and 1000\K\, and revealed
that the BFCO film is a good high-temperature multiferroic with a magnetic phase
transition that we assumed to take place somewhere between 600 and 800\K. Conversion
electron M\"{o}ssbauer spectrum did not reveal magnetic order in the BFCO thin film in
contrast to vibration magnetometry, because the electric field presented in ferroelectric
domains extends the track of emitted electrons and prevents their detection from most the
volume depth of the thin film. Therefore probably only electrons from the thin skin
non-magnetic layer of the film are detected. Further structural, magnetic and dielectric
high-temperature studies on a well B-site-ordered samples are in progress.

\begin{acknowledgments}

The work was supported by the Grant Agency of the Czech Republic (Project No.
202/06/0403) and Grant Agency of Academy of Science of the Czech Republic (Project No.
KJB100100704) and AVOZ10100520. Part of this research was supported by grants from the
Natural Sciences and Engineering Research Council of Canada (NSERC) as well as from the
Fond Qu\'{e}b\'{e}cois de la Recherche sur la Nature et les Technologies (FQRNT). The
authors thank J. \v{S}ebek for valuable discussions as well as J. Petzelt and T.W.
Johnston for critical reading of the manuscript.

\end{acknowledgments}

\end{document}